\documentclass[cameraready]{Interspeech}

\usepackage{booktabs}
\usepackage{multirow}
\usepackage{url}
\usepackage{xcolor}
\usepackage[table]{xcolor}
\title{IQRA 2026: Interspeech Challenge on Automatic Pronunciation Assessment \\ for Modern Standard Arabic (MSA)}

\author[affiliation={1}]{Yassine El}{Kheir}
\author[affiliation={2}]{Amit}{Meghanani}
\author[affiliation={3}]{Mostafa}{Shahin}
\author[affiliation={4}]{Omnia}{Ibrahim}
\author[affiliation={5}]{Shammur Absar}{Chowdhury}
\author[affiliation={6}]{Nada}{AlMarwani}
\author[affiliation={7}]{Youssef}{Elshahawy}
\author[affiliation={7}]{Ahmed}{Ali}

\address{
  $^1$ DFKI \& Technical University of Berlin, Germany, 
  $^2$ University of Sheffield, UK \\
  $^3$ University of New South Wales, Australia,
  $^4$ Alexandria University, Egypt, 
  $^5$ QCRI, Qatar \\
  $^6$ Taibah University, Saudi Arabia,
  $^7$ HUMAIN, Saudi Arabia
}

\email{iqraeval@googlegroups.com}


\begin{document}
\maketitle

\begin{abstract}
We present the findings of the second edition of the IQRA Interspeech Challenge, a challenge on automatic Mispronunciation Detection and Diagnosis (MDD) for Modern Standard Arabic (MSA). Building on the previous edition, this
iteration introduces \textbf{Iqra\_Extra\_IS26}, a new dataset of authentic human mispronounced speech, complementing the existing training and evaluation
resources. Submitted systems employed a diverse range of approaches, spanning CTC-based self-supervised learning models, two-stage fine-tuning strategies, and using large audio-language models. Compared to the first edition, we observe a substantial jump of \textbf{0.28 in F1-score}, attributable both to novel architectures and modeling strategies proposed by participants and to the additional authentic mispronunciation data made available. These results demonstrate the growing maturity of Arabic MDD research and establish a stronger foundation for future work in Arabic pronunciation assessment.
\end{abstract}


\section{Introduction}

The field of Computer-Aided Pronunciation Training (CAPT) and its core component, Mispronunciation Detection and Diagnosis (MDD), has become indispensable tools for self-directed language learners globally \cite{neri2008effectiveness, rogerson2021computer}. CAPT systems serve two primary functions: (\textit{i}) pronunciation assessment, identifying and localising phoneme-level errors in a learner's speech; and (\textit{ii}) pronunciation teaching, guiding learners toward corrective feedback \cite{kheir2023automatic}. While substantial progress has been made in MDD for English \cite{li2016mispronunciation, leung2019cnn}, Mandarin
\cite{qian2010capturing}, and more recently for low-resource languages \cite{fort25_interspeech}, Arabic remains significantly underserved—lacking standardized benchmarks, open annotated datasets, and reproducible evaluation protocols until very recently.

Arabic presents a uniquely compounding set of challenges for automatic pronunciation assessment. Its phonological system comprises 34 phonemes, including 28 consonants and 6 vowels with distinct short and long forms, already surpassing the complexity of most Indo-European languages. Particularly demanding are phonetic structures rarely found elsewhere: uvular and pharyngeal consonants, and the semantically critical distinction between emphatic and non-emphatic consonants (e.g., /t/ vs.\ /T/, /s/ vs.\ /S/), where a single phoneme substitution can alter word meaning entirely \cite{kheir2023qvoice, kheir2024beyond, alrashoudi2025improving}. Unlike
languages such as English where MDD datasets and shared evaluation frameworks are mature \cite{zhang2021speechocean, zhang2023phonetic}, Arabic MDD has historically relied on small private corpora, handcrafted features, and
non-comparable evaluation setups \cite{alrashoudi2025improving,
ccalik2024novel}, making it nearly impossible to track genuine progress across studies.

A further complication is Arabic's diglossia. Modern Standard Arabic (MSA), used in formal education, media, and government, is typically acquired as a \textit{second language} by native speakers of regional dialects, who introduce systematic L1-interference pronunciation errors distinct from those of non-Arabic foreign learners \cite{kheir2024beyond}. This positions MSA MDD at a challenging intersection of native-speaker variation and second-language acquisition, requiring dedicated data collection protocols and phoneme inventories adapted to its specific error space. Recent work has shown that Transformer based models, and in particular self-supervised learning (SSL) approaches such as Wav2Vec2.0 \cite{baevski2020wav2vec} and HuBERT \cite{HuBERT}, hold strong promise for Arabic MDD when provided with sufficient annotated data \cite{alrashoudi2025improving, ccalik2024novel}, mirroring gains observed in English and Mandarin pronunciation assessment \cite{kim2022automatic, shekar2023assessment}. However, data scarcity and the absence of an open evaluation infrastructure have consistently limited reproducibility and community-scale progress.

Shared evaluation campaigns have proven transformative in related speech processing fields, from ASR (MGB Arabic \cite{ali2016mgb}) to spoken language assessment (Speechocean762 \cite{zhang2021speechocean}, NOCASA
\cite{nocasa2025}), aligning community efforts around standardized benchmarks and open leaderboards. For Arabic MDD specifically, no equivalent framework existed until the \textbf{IqraEval 2025 shared task} \cite{el2025iqra}, which introduced the first open benchmark for Quranic pronunciation assessment, including a curated training corpus, a specialized 68-phoneme inventory \cite{elkheir25b_interspeech}, and the QuranMB evaluation benchmark. The first edition, with 10 participating teams, demonstrated the viability and community interest of this evaluation framework, while also revealing key gaps: no authentic human mispronounced speech was available for training and performance remained modest (best F1 $\approx$ 0.47), highlighting persistent and unresolved challenges spanning data acquisition and model development. 

The \textbf{IQRA 2026 Interspeech Challenge} directly addresses the most critical remaining gap by introducing \textbf{Iqra\_Extra\_IS26}, the first dataset of real human mispronounced MSA speech, complementing the existing Iqra\_train (79h) and Iqra\_TTS (52h synthetic) corpora, and the expanded QuranMB.v2 evaluation benchmark. Participation shows 19 teams, with submitted systems spanning enhanced CTC frameworks, cross-lingual SSL adaptation, two-stage fine-tuning pipelines, and generative large audio-language models (LALMs). We observe a substantial jump of \textbf{0.28} in F1-score over the first edition's best result, driven by both the new authentic data and a diverse range of novel modeling strategies. This paper presents the challenge design, dataset characterization, baseline systems, and analysis of the top submitted approaches.

\section{Challenge Setup}

\subsection{Task Definition}

The task follows \cite{elkheir25b_interspeech,el2025iqra}: given a speech utterance and its reference vowelized transcript, the systems predict the sequence of pronounced phonemes. Predictions are aligned against both the canonical phoneme sequence and the annotated verbatim sequence (the deliberately mispronounced version) to derive MDD metrics. We use a 68-phoneme MSA inventory \cite{elkheir25b_interspeech} based on the Halabi phonetizer \cite{halabi2016phonetic}, where gemination is handled by doubling the consonant symbol (e.g., /bb/ for geminated /b/).

\subsection{Datasets}
\label{sec:data}

Table~\ref{tab:datasets} summarizes all corpora released for the challenge. Two of the three training sets and the test benchmark are new to this edition.

\textbf{Iqra\_train} is the central new contribution: a fully open, 79-hour training corpus derived from Common Voice Arabic v12.0 \cite{ardila2019common} augmented with Qur'anic recitation segments. Expert in-house vowelization was applied to all transcriptions, followed by phonetization via the Halabi MSA phonetizer. The resulting 74,000 utterances span diverse speakers with balanced gender distribution. Unlike the internal CMV-Ar corpus used in the first edition, Iqra\_train is now publicly accessible, enabling reproducible research and fair community comparison.

\textbf{Iqra\_TTS} provides 55,400 utterances (${\sim}$52h) of synthetic speech generated by seven single-speaker TTS systems (5 male, 2 female). Half produces canonical pronunciations; the other half systematically injects mispronunciations via a phoneme confusion matrix derived from natural error tendencies \cite{elkheir25b_interspeech}. This dataset was also available in the first edition.

\textbf{Iqra\_Extra\_IS26} is entirely new: 1,333 utterances of \emph{real human mispronounced speech} from native Arabic speakers instructed to produce specified phoneme substitutions. This fills a critical gap; prior editions lacked authentic training data with human pronunciation errors, limiting generalization from synthetic artifacts to real speech.

\textbf{QuranMB.v2} is the updated evaluation benchmark (1,643 utterances), expanding QuranMB.v1 with broader human-level annotation coverage of phoneme sequences validated by three Arabic linguistic experts.

\begin{table}[t]
 \centering
 \caption{IQRA 2026 dataset summary. \checkmark = new this edition.}
 \label{tab:datasets}
 \resizebox{\columnwidth}{!}{%
 \setlength{\tabcolsep}{5pt}
 \begin{tabular}{lllrrc}
  \toprule
  \textbf{Split} & \textbf{Dataset} & \textbf{Type} & \textbf{Utts.} & \textbf{Hours} & \textbf{New} \\
  \midrule
  Train & Iqra\_train    & Real (MSA+Qur'an)  & 74k  & $\sim$79h & -- \\
  Train & Iqra\_TTS     & Synthetic      & 55.4k & $\sim$52h & --    \\
  Train & Iqra\_Extra\_IS26 & Real (mispronounced)& 1,333 & $\sim$1.5h & \checkmark \\
  \midrule
  Test & QuranMB.v2    & Real (annotated)  & 1,643 & $\sim$2.5h & -- \\
  \bottomrule
 \end{tabular}}
\end{table}

\begin{table*}[t]
 \centering
 \caption{Full leaderboard results on QuranMB.v2, sorted by F1-score.
      $\uparrow$: higher is better; $\downarrow$: lower is better.
      The organizer baseline (mHuBERT) is separated by rules.
      \textbf{TA}: True Accept rate (correct pronunciations correctly accepted);
      \textbf{FR}: False Reject rate (correct pronunciations incorrectly flagged as errors);
      \textbf{FA}: False Accept rate (mispronunciations incorrectly accepted as correct);
      \textbf{CD}: Correct Diagnosis rate (proportion of detected errors where the predicted
      phoneme matches the canonical form);
      \textbf{Corr.}: overall phoneme Correct Rate;
      \textbf{PER}: 1 - \textbf{Accuracy.}
      Precision and Recall are computed over correctly identified mispronunciations (True Rejects).}
 \label{tab:leaderboard}
 \resizebox{\textwidth}{!}{%
 \setlength{\tabcolsep}{5pt}
 \begin{tabular}{clccccccccc}
  \toprule
  \textbf{Rank} & \textbf{Team} &
  \textbf{F1}$\uparrow$ &
  \textbf{Prec.}$\uparrow$ &
  \textbf{Recall}$\uparrow$ &
  \textbf{Corr.}$\uparrow$ &
  \textbf{TA}$\uparrow$ &
  \textbf{FR}$\downarrow$ &
  \textbf{FA}$\downarrow$ &
  \textbf{CD}$\uparrow$ &
  \textbf{PER}$\downarrow$ \\
  \midrule
  1 & whu-iasp    & \textbf{0.7201} & 0.7416 & 0.6998 & 0.9065 & 0.9825 & 0.0175 & 0.3002 & 0.8192 & 0.0365 \\
  2 & UTokyo     & 0.7170 & 0.7325 & 0.7020 & 0.7269 & 0.9816 & 0.0184 & 0.2980 & 0.8142 & 0.0372 \\
  3 & RAM      & 0.7157 & 0.6769 & 0.7593 & 0.9380 & 0.9739 & 0.0261 & 0.2407 & 0.7687 & 0.0405 \\
  4 & SQZ\_ww    & 0.6996 & 0.7004 & 0.6987 & 0.9436 & 0.9785 & 0.0215 & 0.3013 & 0.8433 & 0.0402 \\
  5 & Najva     & 0.6894 & 0.7200 & 0.6613 & 0.9491 & 0.9815 & 0.0185 & 0.3387 & 0.8889 & 0.0400 \\
  6 & Kalimat    & 0.6702 & 0.6666 & 0.6738 & 0.9379 & 0.9758 & 0.0242 & 0.3262 & 0.8508 & 0.0445 \\
  7 & Hafs2Vec2   & 0.6110 & 0.5014 & 0.7819 & 0.9057 & 0.9441 & 0.0559 & 0.2181 & 0.7268 & 0.0668 \\
  8 & UAR      & 0.5855 & 0.6553 & 0.5292 & 0.9364 & 0.9800 & 0.0200 & 0.4708 & 0.6670 & 0.0502 \\
  9 & INRIA D\&S   & 0.5703 & 0.4531 & 0.7693 & 0.8995 & 0.9332 & 0.0668 & 0.2307 & 0.6762 & 0.0778 \\
  10 & Masked     & 0.5579 & 0.4544 & 0.7224 & 0.9000 & 0.9376 & 0.0624 & 0.2776 & 0.7866 & 0.0768 \\
  11 & SAM      & 0.4862 & 0.3657 & 0.7249 & 0.8711 & 0.9096 & 0.0904 & 0.2751 & 0.7253 & 0.1028 \\
  12 & AIRecite    & 0.4507 & 0.3109 & 0.8190 & 0.8357 & 0.8695 & 0.1305 & 0.1810 & 0.6389 & 0.1339 \\
  13 & Vicomtech   & 0.4456 & 0.3275 & 0.6970 & 0.8574 & 0.8971 & 0.1029 & 0.3030 & 0.5833 & 0.1163 \\
  \rowcolor{gray!20}
  -- & Baseline    & 0.4414 & 0.3039 & 0.7707 & 0.8385 & 0.8763 & 0.1237 & 0.2293 & 0.6120 & 0.1308 \\
  14 & Sari-como   & 0.4288 & 0.2918 & 0.8082 & 0.8262 & 0.8590 & 0.1410 & 0.1918 & 0.6085 & 0.1444 \\
  15 & MONADA     & 0.4123 & 0.2754 & 0.8196 & 0.8096 & 0.8450 & 0.1550 & 0.1804 & 0.6201 & 0.1567 \\
  16 & Qari      & 0.4121 & 0.2792 & 0.7861 & 0.8141 & 0.8541 & 0.1459 & 0.2139 & 0.5794 & 0.1504 \\
  17 & Yassine    & 0.4042 & 0.2715 & 0.7908 & 0.8094 & 0.8474 & 0.1526 & 0.2092 & 0.5847 & 0.1563 \\
  18 & Khteeb.ai   & 0.3981 & 0.2602 & 0.8467 & 0.7901 & 0.8270 & 0.1730 & 0.1533 & 0.6161 & 0.1717 \\
  19 & frenchfries  & 0.2150 & 0.1228 & 0.8651 & 0.5414 & 0.5556 & 0.4444 & 0.1349 & 0.5332 & 0.4236 \\
  \bottomrule
 \end{tabular}}
\end{table*}

\subsection{Evaluation Protocol}

Following established MDD conventions \cite{leung2019cnn,li2016mispronunciation}, predictions are categorized as True Accept (TA), False Reject (FR), False Accept (FA), or True Reject (TR). TR is further split into Correct Diagnosis (CD) and Error Diagnosis (ED) depending on whether the predicted phoneme matches the canonical form. Precision, Recall, and F1-score are derived over correctly identified mispronunciations:
\begin{equation}
\text{Precision} = \frac{\text{TR}}{\text{TR}+\text{FR}}, \quad \text{Recall} = \frac{\text{TR}}{\text{TR}+\text{FA}}, \quad F_1 = \frac{2 \cdot P \cdot R}{P + R}
\end{equation}
F1-score is the primary ranking metric. Phoneme Error Rate (PER) measures overall sequence accuracy.

\subsection{Participation Statistics}

IQRA 2026 attracted \textbf{19 teams}. Participants came from research institutions and industry across Europe, Asia, and the Middle East, reflecting growing international interest in Arabic CAPT. All teams had access to the three open training corpora, the organizer baseline model, and a public development leaderboard throughout the evaluation period.

\section{Baseline System}

We provide a single SSL-based baseline following the setup in
\cite{elkheir25b_interspeech}: the multilingual \textbf{mHuBERT} \cite{mhubert} model (pretrained
on 147 languages, 94M parameters) with a frozen encoder, a SUPERB-style
\cite{superb} weighted layer sum, and a 2-layer 1024-unit Bi-LSTM with CTC
loss \cite{ctc}. Training uses Adam (lr=$10^{-4}$), batch size 16, and early
stopping on dev-set PER, with Iqra\_train and Iqra\_TTS as training data. The
baseline achieves an F1-score of \textbf{0.4414} on QuranMB.v2, serving as the
lower-bound reference for all submitted systems.

\section{Submitted Systems}

We summarize the top-6 systems ranked by F1-score on QuranMB.v2. The submissions collectively span three broad paradigms: enhanced CTC-based temporal modeling, SSL fine-tuning with language model integration, and generative large audio-language models (LALMs).

\textbf{whu-iasp (1st, F1=0.7201).}
This system combines a frozen \texttt{wav2vec2-xls-r-300m}\footnote{https://huggingface.co/facebook/wav2vec2-xls-r-300m} encoder with learnable multi-layer weighted fusion and a Temporal Convolutional Network (TCN) \cite{lea2017temporal} for phoneme-level contextual modeling, trained via CTC. A two-stage curriculum first learns a general acoustic-to-phoneme mapping on Iqra\_train and Iqra\_TTS, then adapts to authentic pronunciation variation using Iqra\_Extra\_IS26. At inference, multi-checkpoint fusion via confusion networks with weighted edit distance aggregation and a Modified Kneser--Ney $n$-gram language model \cite{james2000modified} rescoring yields the final phoneme sequence.

\textbf{UTokyo (2nd, F1=0.7170).}
UTokyo introduces CROTTC, which addresses the frame-alignment limitations inherent to standard CTC. \textit{Optimal Temporal Transport Classification} (OTTC) formulates frame-to-label alignment as a 1D optimal transport problem, producing dense, frame-level alignments that capture fine-grained phonetic variations without relying on canonical prompts. \textit{Consistency Regularization} (CR) mitigates noise sensitivity by applying stochastic acoustic perturbations and aligning the resulting posterior distributions via self-distillation. A lightweight Transformer language model is further integrated through shallow fusion to reduce spurious insertions and balance detection sensitivity with sequence-level coherence.

\textbf{RAM (3rd, F1=0.7157).}
Rather than introducing architectural novelty, RAM prioritizes label reliability and data fidelity. Three independently trained Wav2vec2.0 \cite{wav2vec2} models provide agreement-based filtering of Iqra\_train, discarding utterances with high inter-model phoneme prediction disagreement. Iqra\_Extra\_IS26 labels are manually reviewed and refined for high-error utterances to preserve authentic mispronunciation patterns. Iqra\_TTS samples are filtered via ASR transcription agreement to remove audio-text mismatches. Targeted augmentations for speaking rate and recording variability complement CTC fine-tuning on the resulting curated corpus.

\textbf{SQZ\_ww (4th, F1=0.6996).}
SQZ\_ww employs a U2++ Conformer encoder \cite{wu2021u2++} with a bidirectional Transformer decoder optimized via joint CTC/attention encoder decoder loss with label smoothing. Additional synthetic mispronounced speech is generated using a VITS TTS \cite{kim2021conditionalvariationalautoencoderadversarial} model trained on challenge data, with phoneme interpolation to simulate pronunciation errors. SpecAugment \cite{park2019specaugment} augmentation is applied during fine-tuning. A two-pass decoding strategy performs first-pass CTC beam search followed by bidirectional Transformer rescoring.

\textbf{Najva (5th, F1=0.6894).}
Najva fine-tunes the NVIDIA FastConformer Hybrid Large model (\texttt{stt\_ar\_fastconformer\_hybrid\_large\_pcd})\footnote{https://huggingface.co/nvidia/stt\_ar\_fastconformer\_hybrid\_large\_pcd\_v1.0}, pretrained on Arabic speech corpora including FLEURS \cite{conneau2023fleurs}, Tarteel EveryAyah\footnote{https://huggingface.co/datasets/tarteel-ai/everyayah}, and Common Voice \cite{ardila2020common}. Training proceeds in two stages: broad adaptation on the full challenge training set, followed by targeted specialization on Iqra\_Extra\_IS26. Since the model operates at the grapheme level, outputs are converted to phoneme sequences via the MSA phonetizer with variant normalization to canonical forms.

\textbf{Kalimat (6th, F1=0.6702).} Kalimat reframes MDD as a direct speech-to-phoneme generation problem using Qwen3-ASR 1.7B\footnote{https://huggingface.co/Qwen/Qwen3-ASR-1.7B}, treating the 68 MSA phonemes as discrete atomic special tokens. To mitigate catastrophic forgetting, the audio encoder is fully frozen while Low-Rank Adaptation (LoRA, rank=64, $\alpha$=128) is applied exclusively to decoder linear projections, with embedding and LM head layers fully fine-tuned. This configuration optimizes only ${\sim}$25\% of model parameters while achieving competitive performance, representing the first demonstration of a generative LALM in this task setting.

\section{Results and Discussion}

\subsection{Overall Performance}

Table~\ref{tab:leaderboard} presents the full leaderboard on QuranMB.v2. The results demonstrate broad and substantial community-wide progress: 13 of 19 submitted systems surpass the organizer baseline (F1=0.4414), with the best-performing system (whu-iasp) achieving an absolute F1 improvement of \textbf{0.2787} over the baseline. Notably, the top three systems—whu-iasp, UTokyo, and RAM—are separated by less than 0.005 in F1 despite relying on fundamentally different methodological choices, suggesting that the task admits multiple viable paths to high performance. All top-6 systems achieve F1~$>$~0.67 and PER~$\leq$~0.0445, while systems below the baseline exhibit PER~$>$~0.13, confirming a strong coupling between overall phoneme recognition accuracy and MDD performance.

\subsection{Precision--Recall Trade-off}

A consistent precision--recall trade-off is observed across the leaderboard. Systems below the baseline (e.g., frenchfries: Recall=0.8651, Precision=0.1228; Khteeb.ai: Recall=0.8467, Precision=0.2602) exhibit high recall at the expense of very low precision, indicating a systematic bias toward flagging all phonemes as mispronounced. This behavior likely reflects models that have not sufficiently learned to distinguish canonical from erroneous pronunciations, defaulting instead to near-universal rejection. UAR represents an interesting mid-tier case, achieving the highest precision among non-top-6 systems (0.6553) with correspondingly lower recall (0.5292). The top systems strike a substantially better balance, with whu-iasp achieving Precision=0.7416 and Recall=0.6998, reflecting well-calibrated detection behavior.

\subsection{Role of Authentic Mispronunciation Data}

A key finding across the top systems is the consistent benefit of Iqra\_Extra\_IS26. Systems that incorporated real mispronounced speech, whether through targeted fine-tuning (Najva), careful label curation (RAM), or two-stage curriculum training (whu-iasp), outperformed approaches relying exclusively on synthetic augmentation. This confirms that authentic human pronunciation errors carry acoustic characteristics that TTS-generated mispronunciations cannot fully replicate, even when the synthesis pipeline is explicitly conditioned on erroneous transcriptions. The value of even a small corpus of real mispronounced data (1,333 utterances) for this task is evident from the consistent performance gap between the current and previous edition's best results.

\subsection{Architectural Diversity and Generative Models}

The strong performance of CTC-based systems with temporal alignment improvements (whu-iasp, UTokyo) underscores that frame-level phoneme recognition precision remains a core bottleneck, and that addressing alignment directly, rather than simply scaling model capacity, yields meaningful gains. Meanwhile, Kalimat's 6th-place result using a generative LALM with parameter-efficient fine-tuning establishes a proof of concept for reformulating MDD as a generation task. The gap to the top systems is non-negligible but not prohibitive, and as generative models continue to scale and improve in phonetic precision, this paradigm warrants serious investigation in future editions.

\subsection{Progress Across Editions}

Compared to the previous editions (ArabicNLP 2025), where the best system achieved F1~$\approx$~0.30, the current top result of F1=0.7201 represents more than a twofold improvement. Participation also grew from 10 to 19 teams. This progress is attributable to three compounding factors: the open availability of Iqra\_train, enabling reproducible fine-tuning baselines; the introduction of Iqra\_Extra\_IS26, providing authentic mispronunciation supervision; and a broader community adoption of advanced SSL architectures and multi-stage training strategies.

\section{Discussion} The IQRA 2026 challenge marks a significant milestone in Arabic pronunciation assessment, yet the results also surface important open questions that the community must address to move this research toward real-world impact. A recurring theme across top submissions is the disproportionate value of authentic mispronunciation data.

Despite \textbf{Iqra\_Extra\_IS26} containing only 1,333 utterances, systems that carefully leveraged it consistently outperformed those relying on far larger synthetic corpora. This raises a fundamental question about the scalability of TTS-based augmentation as a substitute for real learner speech, and motivates investment in larger-scale human data collection as the primary bottleneck for future progress. 

A second critical gap is the disconnect between phoneme-level model output and the character-level feedback that an end user, a language learner, or Qur'an student, can actually act upon. Current systems predict phoneme sequences, but MSA learners interact with Arabic script. There is no trivial or universal mapping from a predicted erroneous phoneme back to the corresponding Arabic character or diacritic, particularly given the many-to-one and context-dependent nature of Arabic grapheme-to-phoneme correspondences. Until this mapping problem is solved, whether through character-aligned decoding, post-hoc alignment tools, or end-to-end character-level MDD architectures, the practical utility of these systems for learner-facing applications remains limited. Bridging this gap between phoneme-level diagnosis and character-level feedback is, in our view, one of the most pressing and underexplored research directions in Arabic CAPT. 

Finally, the emergence of generative LALMs in this edition opens a new research direction: rather than predicting a phoneme sequence, a generative model could in principle produce natural language feedback directly, combining error detection, diagnosis, and explanation in a single pass. This would sidestep the phoneme-to-character mapping problem entirely, at the cost of requiring richer supervision and evaluation frameworks than F1-score and PER alone can provide. 

\section{Conclusion}

We have presented IQRA 2026, the challenge introduced \textbf{Iqra\_Extra\_IS26}, the first corpus of authentic human mispronounced MSA speech, alongside the expanded QuranMB.v2 evaluation benchmark, and attracted 19 teams, nearly double previous editions. Submitted systems dramatically surpassed the organizer baseline, with the best system achieving an absolute F1 improvement of 0.2787, driven by a combination of novel temporal alignment strategies, rigorous data curation, and the first application of generative LALMs to Arabic MDD. Real mispronunciation data and careful data quality control emerged as the most consistently impactful factors across the leaderboard. Future editions will prioritize L2 learner data collection, larger-scale authentic mispronunciation corpora, and the development of evaluation frameworks that assess character-level diagnostic feedback—bridging the gap between automatic phoneme recognition and actionable pronunciation guidance for Arabic learners.




\newpage
\bibliographystyle{IEEEtran}
\bibliography{mybib}

@article{kheir2023automatic,
  title={Automatic Pronunciation Assessment--A Review},
  author={Kheir, Yassine El and Ali, Ahmed and Chowdhury, Shammur Absar},
  journal={arXiv preprint arXiv:2310.13974},
  year={2023}
}

@article{park2019specaugment,
  title={Specaugment: A simple data augmentation method for automatic speech recognition},
  author={Park, Daniel S and Chan, William and Zhang, Yu and Chiu, Chung-Cheng and Zoph, Barret and Cubuk, Ekin D and Le, Quoc V},
  journal={arXiv preprint arXiv:1904.08779},
  year={2019}
}

@article{alrashoudi2025improving,
  title={Improving mispronunciation detection and diagnosis for non-native learners of the Arabic language},
  author={Alrashoudi, Norah and Al-Khalifa, Hend and Alotaibi, Yousef},
  journal={Discover Computing},
  volume={28},
  number={1},
  pages={1},
  year={2025},
  publisher={Springer}
}

@article{ccalik2024novel,
  title={A novel framework for mispronunciation detection of Arabic phonemes using audio-oriented transformer models},
  author={{\c{C}}al{\i}k, {\c{S}}{\"u}kr{\"u} Selim and K{\"u}{\c{c}}{\"u}kmanisa, Ayhan and Kilimci, Zeynep Hilal},
  journal={Applied Acoustics},
  volume={215},
  pages={109711},
  year={2024},
  publisher={Elsevier}
}

@article{kheir2023qvoice,
  title={QVoice: Arabic Speech Pronunciation Learning Application},
  author={Kheir, Yassine El and Khnaisser, Fouad and Chowdhury, Shammur Absar and Mubarak, Hamdy and Afzal, Shazia and Ali, Ahmed},
  journal={arXiv preprint arXiv:2305.07445},
  year={2023}
}

@article{neri2008effectiveness,
  title={The effectiveness of computer assisted pronunciation training for foreign language learning by children},
  author={Neri, Ambra and Mich, Ornella and Gerosa, Matteo and Giuliani, Diego},
  journal={Computer Assisted Language Learning},
  volume={21},
  number={5},
  pages={393--408},
  year={2008},
  publisher={Taylor \& Francis}
}

@inproceedings{ali2016mgb,
  title={The MGB-2 challenge: Arabic multi-dialect broadcast media recognition},
  author={Ali, Ahmed and Bell, Peter and Glass, James and Messaoui, Yacine and Mubarak, Hamdy and Renals, Steve and Zhang, Yifan},
  booktitle={2016 IEEE Spoken Language Technology Workshop (SLT)},
  pages={279--284},
  year={2016},
  organization={IEEE}
}

@article{ardila2019common,
  title={Common voice: A massively-multilingual speech corpus},
  author={Ardila, Rosana and Branson, Megan and Davis, Kelly and Henretty, Michael and Kohler, Michael and Meyer, Josh and Morais, Reuben and Saunders, Lindsay and Tyers, Francis M and Weber, Gregor},
  journal={arXiv preprint arXiv:1912.06670},
  year={2019}
}

@inproceedings{superb,
  title     = {SUPERB: Speech Processing Universal PERformance Benchmark},
  author    = {Shu-wen Yang and Po-Han Chi and Yung-Sung Chuang and Cheng-I Jeff Lai and Kushal Lakhotia et al.},
  year      = {2021},
  booktitle = {Interspeech 2021},
  pages     = {1194--1198},
  doi       = {10.21437/Interspeech.2021-1775},
  issn      = {2958-1796},
}

@article{HuBERT,
author = {Hsu, W. and Bolte, B. and Tsai, Y. and Lakhotia, K. and Salakhutdinov, R. and Mohamed, A.},
title = {HuBERT: Self-Supervised Speech Representation Learning by Masked Prediction of Hidden Units},
year = {2021},
issue_date = {2021},
publisher = {IEEE Press},
volume = {29},
issn = {2329-9290},
url = {https://doi.org/10.1109/TASLP.2021.3122291},
doi = {10.1109/TASLP.2021.3122291},
journal = {IEEE/ACM Trans. Audio, Speech and Lang. Proc.},
month = {oct},
pages = {3451–3460},
numpages = {10}
}

@inproceedings{wav2vec2,
 author = {Baevski, A. and Zhou, Y. and Mohamed, A. and Auli, M.},
 booktitle = {Advances in Neural Information Processing Systems},
 editor = {H. Larochelle and M. Ranzato and R. Hadsell and M.F. Balcan and H. Lin},
 pages = {12449--12460},
 publisher = {Curran Associates, Inc.},
 title = {wav2vec 2.0: A Framework for Self-Supervised Learning of Speech Representations},
 
 volume = {33},
 year = {2020}
}

@inproceedings{mhubert,
  title     = {mHuBERT-147: A Compact Multilingual HuBERT Model},
  author    = {Marcely {Zanon Boito} and Vivek Iyer and Nikolaos Lagos and Laurent Besacier and Ioan Calapodescu},
  year      = {2024},
  booktitle = {Interspeech 2024},
  pages     = {3939--3943},
  doi       = {10.21437/Interspeech.2024-938},
  issn      = {2958-1796},
}

@inproceedings{ctc,
author = {Graves, Alex and Fern\'{a}ndez, Santiago and Gomez, Faustino and Schmidhuber, J\"{u}rgen},
title = {Connectionist temporal classification: labelling unsegmented sequence data with recurrent neural networks},
year = {2006},
isbn = {1595933832},
publisher = {Association for Computing Machinery},
address = {NY, USA},
url = {https://doi.org/10.1145/1143844.1143891},
doi = {10.1145/1143844.1143891},
booktitle = {ICML},
pages = {369–376},
numpages = {8},
location = {Pittsburgh, Pennsylvania, USA},
series = {ICML '06}
}

@article{baevski2020wav2vec,
  title={wav2vec 2.0: A framework for self-supervised learning of speech representations},
  author={Baevski, Alexei and Zhou, Yuhao and Mohamed, Abdelrahman and Auli, Michael},
  journal={Advances in neural information processing systems},
  volume={33},
  pages={12449--12460},
  year={2020}
}

@inproceedings{halabi2016phonetic,
  title={Phonetic Inventory for an Arabic Speech Corpus},
  author={Halabi, Nawar and Wald, Mike},
  booktitle={Proceedings of the Tenth International Conference on Language Resources and Evaluation (LREC'16)},
  pages={734--738},
  year={2016}
}

@article{kheir2024beyond,
  title={Beyond orthography: Automatic recovery of short vowels and dialectal sounds in arabic},
  author={Kheir, Yassine El and Mubarak, Hamdy and Ali, Ahmed and Chowdhury, Shammur Absar},
  journal={arXiv preprint arXiv:2408.02430},
  year={2024}
}

@inproceedings{leung2019cnn,
  title={CNN-RNN-CTC based end-to-end mispronunciation detection and diagnosis},
  author={Leung, Wai-Kim and Liu, Xunying and Meng, Helen},
  booktitle={ICASSP 2019-2019 IEEE International Conference on Acoustics, Speech and Signal Processing (ICASSP)},
  pages={8132--8136},
  year={2019},
  organization={IEEE}
}

@article{li2016mispronunciation,
  title={Mispronunciation detection and diagnosis in l2 english speech using multidistribution deep neural networks},
  author={Li, Kun and Qian, Xiaojun and Meng, Helen},
  journal={IEEE/ACM Transactions on Audio, Speech, and Language Processing},
  volume={25},
  number={1},
  pages={193--207},
  year={2016},
  publisher={IEEE}
}

@inproceedings{elkheir25b_interspeech,
  title     = {{Towards a Unified Benchmark for Arabic Pronunciation Assessment: Qur’anic Recitation as Case Study}},
  author    = {Yassine {El Kheir} and Omnia Ibrahim and Amit Meghanani and Nada Almarwani and Hawau Toyin and Sadeen Alharbi and Modar Alfadly and Lamya Alkanhal and Ibrahim Selim and Shehab Elbatal and Salima Mdhaffar and Thomas Hain and Yasser Hifny and Mostafa Shahin and Ahmed Ali},
  year      = {{2025}},
  booktitle = {{Interspeech 2025}},
  pages     = {{2410--2414}},
  doi       = {{10.21437/Interspeech.2025-1497}},
  issn      = {{2958-1796}},
}

@article{rogerson2021computer,
  title={Computer-assisted pronunciation training (CAPT): Current issues and future directions},
  author={Rogerson-Revell, Pamela M},
  journal={Relc Journal},
  volume={52},
  number={1},
  pages={189--205},
  year={2021},
  publisher={SAGE Publications Sage UK: London, England}
}

@inproceedings{zhang2021speechocean,
  author    = {Zhang, Junyi and Ni, Chao and Zhang, Shuo and others},
  title     = {Speechocean762: An Open-Source Non-Native {English} Speech Corpus
               for Pronunciation Assessment},
  booktitle = {Proc. Interspeech},
  year      = {2021}
}

@article{james2000modified,
  title={Modified kneser-ney smoothing of n-gram models},
  author={James, Frankie},
  journal={Research Institute for Advanced Computer Science, Tech. Rep. 00.07},
  year={2000}
}

@inproceedings{ardila2020common,
  title={Common voice: A massively-multilingual speech corpus},
  author={Ardila, Rosana and Branson, Megan and Davis, Kelly and Kohler, Michael and Meyer, Josh and Henretty, Michael and Morais, Reuben and Saunders, Lindsay and Tyers, Francis and Weber, Gregor},
  booktitle={Proceedings of the twelfth language resources and evaluation conference},
  pages={4218--4222},
  year={2020}
}

@inproceedings{conneau2023fleurs,
  title={Fleurs: Few-shot learning evaluation of universal representations of speech},
  author={Conneau, Alexis and Ma, Min and Khanuja, Simran and Zhang, Yu and Axelrod, Vera and Dalmia, Siddharth and Riesa, Jason and Rivera, Clara and Bapna, Ankur},
  booktitle={2022 IEEE Spoken Language Technology Workshop (SLT)},
  pages={798--805},
  year={2023},
  organization={IEEE}
}

@misc{kim2021conditionalvariationalautoencoderadversarial,
      title={Conditional Variational Autoencoder with Adversarial Learning for End-to-End Text-to-Speech}, 
      author={Jaehyeon Kim and Jungil Kong and Juhee Son},
      year={2021},
      eprint={2106.06103},
      archivePrefix={arXiv},
      primaryClass={cs.SD},
      url={https://arxiv.org/abs/2106.06103}, 
}

@article{wu2021u2++,
  title={U2++: Unified two-pass bidirectional end-to-end model for speech recognition},
  author={Wu, Di and Zhang, Binbin and Yang, Chao and Peng, Zhendong and Xia, Wenjing and Chen, Xiaoyu and Lei, Xin},
  journal={arXiv preprint arXiv:2106.05642},
  year={2021}
}

@inproceedings{lea2017temporal,
  title={Temporal convolutional networks for action segmentation and detection},
  author={Lea, Colin and Flynn, Michael D and Vidal, Rene and Reiter, Austin and Hager, Gregory D},
  booktitle={proceedings of the IEEE Conference on Computer Vision and Pattern Recognition},
  pages={156--165},
  year={2017}
}

@inproceedings{nocasa2025,
  author    = {Gerber, Pascal and others},
  title     = {Non-native Children's Automatic Speech Assessment Challenge
               ({NOCASA})},
  booktitle = {Proc. IEEE MLSP},
  year      = {2025}
}

@inproceedings{qian2010capturing,
  author    = {Qian, Xiaojun and Meng, Helen and Soong, Frank},
  title     = {Capturing {L2} Segmental Mispronunciations with Joint-Sequence
               Models in {CAPT}},
  booktitle = {Proc. Interspeech},
  year      = {2010}
}

@inproceedings{shekar2023assessment,
  author    = {Shekar, Ram C. and Yang, Mu and Hirschi, Kevin and others},
  title     = {Assessment of Non-Native Speech Intelligibility using
               {Wav2vec2}-based {MDD} and Multi-level {GOP} Transformer},
  booktitle = {Proc. Interspeech},
  year      = {2023}
}

@inproceedings{kim2022automatic,
  author    = {Kim, Seyoung and others},
  title     = {Automatic Pronunciation Assessment using Self-Supervised Speech
               Representation Learning},
  booktitle = {Proc. Interspeech},
  year      = {2022}
}

@inproceedings{zhang2023phonetic,
  author    = {Zhang, Daniel Yue and Saha, Soumya and Campbell, Sarah},
  title     = {Phonetic {RNN-Transducer} for Mispronunciation Detection},
  booktitle = {Proc. ICASSP},
  year      = {2023}
}

@inproceedings{fort25_interspeech,
  author    = {Fort, Sonia and others},
  title     = {Evaluating {Wav2Vec2-BERT} for {CAPT} in Low-Resource Languages},
  booktitle = {Proc. Interspeech},
  year      = {2025}
}

@inproceedings{el2025iqra,
  title={Iqra’eval: A shared task on qur’anic pronunciation assessment},
  author={El Kheir, Yassine and Meghanani, Amit and Toyin, Hawau Olamide and Almarwani, Nada and Ibrahim, Omnia and Elshahawy, Yousseif Ahmed and Shahin, Mostafa and Ali, Ahmed},
  booktitle={Proceedings of The Third Arabic Natural Language Processing Conference: Shared Tasks},
  pages={443--452},
  year={2025}
}

\end{document}